\documentclass{article}
\usepackage[dvips]{epsfig}
\usepackage{amsmath}
\usepackage{amssymb}
\usepackage{alltt,array}
\renewcommand{\vec}[1]{{\bf #1}}
\newcommand{\angen}{$\,\frac{\mbox{kJ}}{\mbox{mol}\,\mbox{rad}^{2}}$}
\begin{document}
\title{Automatic Parameterization of Force Fields for Liquids by
  Simplex Optimization}
\author{Roland Faller, \and Heiko Schmitz, \and Oliver Biermann, \and Florian
  M\"uller-Plathe \\
\small Max-Planck Institut f\"ur Polymerforschung, Ackermannweg 10, 
D-55128 Mainz}
\maketitle 
\extrarowheight 0.1cm
\noindent {\bf Abstract} We demonstrate an automatic method of force field
development for molecular simulations. Parameter tuning is taken as an
optimization problem in many dimensions. The parameters are automatically
adapted to reproduce known experimental data such as the density and the heat
of vaporization. Our method is more systematic than guessing parameters and,
at the same time saves human labour in parameterization. It was successfully
applied to several molecular liquids: As a test, force fields for
2-methylpentane, tetrahydrofurane, cyclohexene and cyclohexane were
developed. 

\vspace{1ex}


%\vspace{1ex}

%\noindent {\bf For submission to \it J Comput Chem}
%
\section{Introduction}
In atomistic molecular dynamics simulations, one of the central problems is
the choice of the proper parameters for modeling the desired system. There is
a variety of approaches to this problem. Ab initio quantum chemistry would be
an ideal tool for this purpose if it were able to handle interactions of big
molecules in reasonable time. The standard solution, however, is quite
pragmatic.  One either chooses a force field that reproduces certain
experimental data or one takes standard values for the different atoms. Hence,
force field design is either a cumbersome trial-and-error procedure or relies
heavily on the transferability of parameters.

There are attempts to make the computer do this job, e.g. force field
development by weak coupling \cite{njo95,berweger95}.
%changed for revision
However, that procedure relies on the requirements that one force field
parameter dominates the behavior of one property and that their relationship
is monotonic. As, in more complex force fields, one property may be influenced 
significantly by several parameters, a more general multidimensional
optimization algorithm is needed.
%change end
In our approach, we consider the experimentally measured properties as
multi-dimensional functions of the parameters. Then we use the well-known
simplex algorithm \cite{press92} to find the optimum parameter set.
\section{Algorithm and Implementation}
\subsection{Simplex algorithm}
The simplex method is a well-known algorithm for minimization in many
dimensions \cite{press92}. It is not constrained by conditions like
% convexity added
monotonicity, convexity or differentiability of the function being
optimized. It minimizes any single-valued function of an arbitrary number of
variables. Additionally, it is very robust in finding a local optimum. Its
main drawback is the large number of necessary function evaluations, i.e in
our case MD simulation runs, which are quite time consuming. In the following
we briefly summarize the simplex algorithm used in this work.

\begin{figure}
  \epsfxsize 12cm
  \epsfbox{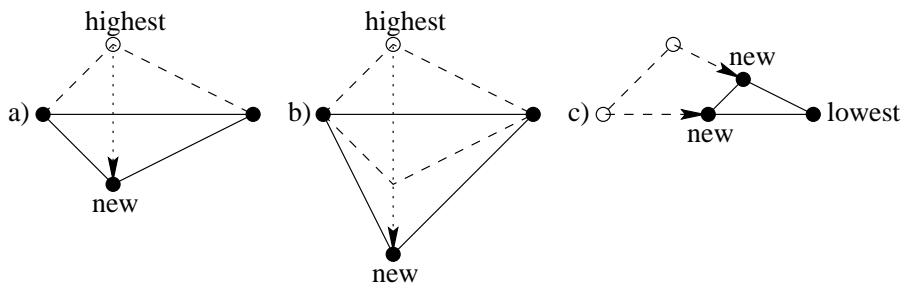}
  \caption{ Transformations of the simplex used during the algorithm: 
    a) reflection, b) expansion, c) contraction.}
  \label{fig:sketch}
\end{figure}

A simplex (a `` $d$-dimensional distorted tetrahedron'') is a set of $d+1$
points in the $d$-dimensional parameter space.  It is transformed
geometrically depending upon the ``quality'' of the function values. There are
three geometric transformations in the algorithm.  
\begin{enumerate}
\item In a reflection, the point $ \vec{x_{i}}$ with the highest
  function value is reflected through the hyper-plane defined by the other
  points (see figure \ref{fig:sketch}a). 
  \begin{equation}
    \vec{x_{i}^{\prime}}= \frac{2}{d}\sum_{j=1}^{d+1}
    \vec{x_{j}}-
    \left(\frac{2}{d}+1\right)\vec{x_{i}}
  \end{equation} 
\item An expansion by the factor $\lambda$ is a linear transformation of one
  point along the normal of the hyper-plane defined by the others
  (fig. \ref{fig:sketch}b).  
  \begin{equation}
    \vec{x_{i}^{\prime}}= \frac{1-\lambda}{d}\sum_{j=1}^{d+1}
    \vec{x_{j}}-
    \left(\frac{1-\lambda}{d}+1\right)\vec{x_{i}}
  \end{equation} 
  Thus, a reflection is just the special case $\lambda=-1$.
\item A ($d$-dimensional) contraction is a linear transformation of all
  but one point $\vec{x_{j}}$ towards the lowest point
  (fig. \ref{fig:sketch}c). Contractions by a factor of 2 are applied. 
  \begin{equation}
    \vec{x_{i}^{\prime}}=
    \frac{1}{2}(\vec{x_{i}}+\vec{x_{j}}),\forall i\ne j 
  \end{equation}
\end{enumerate}
The algorithm runs iteratively. Each iteration starts with a reflection of
the highest point. Depending on the function value at the new point, an
expansion or a contraction is performed. If the new point is better than the
best point an additional expansion with the factor $\lambda=2$ (i.e. the
distance to the hyper-plane of the others is doubled) is applied to explore
further into this ``promising'' direction. If the new point point is very far
away from the minimum (i.e. worse than the second worst point up to now) an
expansion with $\lambda=0.5$ is applied. If this resulting point is still very
bad (in the above sense) a contraction around the best point is
performed. Then the next iteration, starting again with a reflection,
follows. 
\subsection{The target function evaluation}
As the algorithm only knows about scalar functions in ${\cal R}^{d}$, we have
to construct a single-valued function $f_{target}(p_{1},\ldots,p_{d})$ of our
force field parameters $p_{1},\ldots,p_{d}$. The function to be minimized
should indicate the deviation of physical properties of the simulated model
system from the real system as observed in experiments. Typically, one chooses
a set of physical properties $\{P_{i}\}$, which are well characterized
experimentally and converge rapidly in simulations. A natural choice for
$f_{target}$ is the square root of the weighted sum of relative squared
deviations
\begin{equation}
  f_{target}(\{p_{n}\}) = \left(\sum_{i} w_{i}\left(1-\frac{P_{i}(\{p_{n}\})}
    {P_{i,target}}\right)^{2}\right)^{1/2},
\end{equation}
where $P_{i,target}$ is the experimental value of property $P_{i}$. 
%referee
The square root is chosen because it comes steeper to the minimum.
The weights $w_{i}$ account for the fact that some property may
be easier to reproduce than others. Thus, the algorithm can be forced to
focus stronger on the difficult properties. Typically, the density 
$\rho$ is easier reproduced than the enthalpy of vaporization $\Delta
H_{vap}$, which are the two properties we optimize our force fields
against. They converge rapidly and experimental data is readily available for
many fluids (see e.g. \cite{landolt93,CRC}).

If the number of parameters to be optimized is about 2 to 4 the flexibility to
fit the data is normally sufficient and the computational time is still
manageable. If there are more target properties it may be necessary to
increase the dimensionality of the optimization space at the cost of more
computer time.
 
In the beginning, a simplex of parameter sets has to be constructed by the
user. These data may be guessed from parameters for similar compounds or from
standard force fields \cite{amber95,gromos96,sun98a}. Furthermore, a starting
configuration of the system is needed which should be close to the supposed
real state. That means that geometry and density should be almost correct. The
starting configuration is relaxed some picoseconds with a guessed force field
in order to obtain a proper liquid structure. The target function for the
initial parameter sets is first evaluated before the simplex algorithm starts.
\subsection{Parameters to optimize}
Since the dimensionality of parameter space is limited, we have to decide
which parameters of the force field we want to optimize. This number is mainly
limited by the available computing resources. 

Typically, a Lennard-Jones potential is used to model the non-bonded
interactions. 
\begin{equation}
  V_{LJ}=4\epsilon\left(\left(\frac{\sigma}{r}\right)^{12}-
    \left(\frac{\sigma}{r}\right)^{6}\right). \label{eq:LJ}
\end{equation}
The density $\rho$ depends quite strongly on the Lennard-Jones radius $\sigma$
whereas the enthalpy of vaporization $\Delta H_{vap}$ depends stronger on
$\epsilon$. It is recommended to optimize non-bonded interaction parameters or
charges and not the molecular geometry, because of simulational stability. The
fact, that the geometry is mostly quite well known, supports this
choice. There are several experimental methods to determine geometries,
e.g. x-ray or neutron diffraction in the crystal or electron or microwave
diffraction in the gas phase. Ab initio quantum chemistry, too, gives
molecular structures with useful accuracy. These geometries can, in most
cases, be used for the liquid phase as well. Hence, we did not try the
algorithm on geometry optimization although this may be possible in
principle. Our simulations focused on the liquid phase, whose macroscopic
properties depend only weakly on internal force field parameters. Therefore,
the force field parameters for angles and dihedral angles may be adopted from
similar force fields.
\subsection{Equilibration}
A MD run can produce reliable results only if the system has been
equilibrated. Therefore, we need a scheme to test for equilibration which has
to fulfill several requirements: It has to reject reliably non-equilibrated
configurations because otherwise all following results are meaningless. It has
to work fully automatic inside the overall algorithm, and it has to
equilibrate as fast as possible in order not to waste resources.

If the force field parameters (i.e. the Hamiltonian) of a simulation change
between iterations, like in our case, a configuration equilibrated with respect
to the old parameters is no longer equilibrated with respect to the new
ones. Hence, after each change of parameters, i.e. in each step of the simplex
algorithm we have to re-equilibrate with respect to the actual parameters. In
order to do this, we take as the starting configuration the final
configuration from a simulation with a parameter set, which is close to the
new one. As ``distance'' in parameter space we define the sum of squared
deviations 
\begin{equation}
  |\{p^{(new)}\}-\{p^{(old)}\}|^{2} :=
  \sum_{i=1}^{n}(p^{(new)}_{i}-p^{(old)}_{i})^{2}. 
\end{equation}
If, for some reason, the equilibration did not converge for that set or some
other problem occurred a standard configuration is used.

Using the configuration selected in this way we start a number of successive
equilibration runs (typical length 50ps with 1fs timestep). These runs are
analyzed for equilibration until they are either accepted or a maximum number
of runs (in our case 10) is exceeded. In the latter case, the parameters are
considered not useful and the target function $f_{target}$ is set to an
arbitrary high value in order to indicate the failure. 

How does the automatic determination of equilibration work? To our knowledge,
there is no strict criterion for equilibration. The standard procedure is to
inspect visually the time development of a typical quantity (like the density
for low molecular weight liquids). Then one decides if it ``settled'' to
stochastic oscillations around a converged mean value. In our case, we use the
following test: The time series of the density is cut into 3 to 5 intervals,
for each of which the mean and the standard error are calculated. If all these
averages agree within their errors the configuration is considered
equilibrated. In comparison with the ``human eye''-method, this method proved
to be rather strict. However, this is necessary because we cannot accept
non-equilibrated configurations which would mislead the simplex algorithm. The
equilibration scheme worked well and led on average to an equilibrated
configuration in about 3 to 4 runs. Naturally, the number of runs decreases
during the optimization because the changes in the parameters get less
drastic. We also checked a second equilibration test where the last third of
the simulation was fitted by linear regression. If the slope is zero within
its error the configuration is assumed equilibrated. The outcomes of the two
tests differed only slightly.

Only very few parameter sets (less than 10\%) had to be discarded due to
non-equilibration. Even fewer led to instabilities in the simulation. 
\subsection{Convergence criterion}
The simplex algorithm finishes if the target function falls below a given
threshold $l$ which is usually set to about 1\% (i.e. $f_{target}<l\approx
0.01$). If this is achieved the parameters are deemed to be satisfactory. It
does not make sense to reproduce experimental data more closely because the
typical simulation error limits the reliability anyway. In addition, the target
values themselves carry some uncertainty. 

If the desired accuracy $l$ is not achieved and the simplex ends up in a local
minimum the algorithm is aborted. Therefore, the highest and lowest value of
the target function in the actual simplex are compared. Hence, if 
\begin{equation}
  \mbox{max}(f_{target})-\mbox{min}(f_{target}) < \delta f \approx 0.001
\end{equation}
is achieved further optimization makes no sense. In this case, either the
number of parameters is too small to reproduce the desired number of
properties (overdetermination) or the appropriate parameter values are far off
the initial guess. We note that other convergence and abortion criteria are
possible, for example based on the size of the simplex. However, ours have
proven to work well in practice. 
\subsection{Implementation}
The parts of the algorithm were implemented in different programming
languages. The backbone is a {\it tcsh} script which calls all auxiliary
programs and controls the overall flow of the procedure. It uses standard UNIX
utilities like {\it awk}. The routine for producing a new topology from a set
of parameters is a {\it PERL} script whereas the programs for calculating the
distance in parameter space and the determination of equilibration are
implemented in C++. Several programs from the {\it YASP} simulation package
\cite{yasp} are used: the MD program itself as well as the utilities for
calculating enthalpy of vaporization and density.  Any program or utility may
be easily exchanged without affecting the overall structure, e. g. for using
another MD program or a different equilibration scheme.

\begin{figure}
  \epsfbox{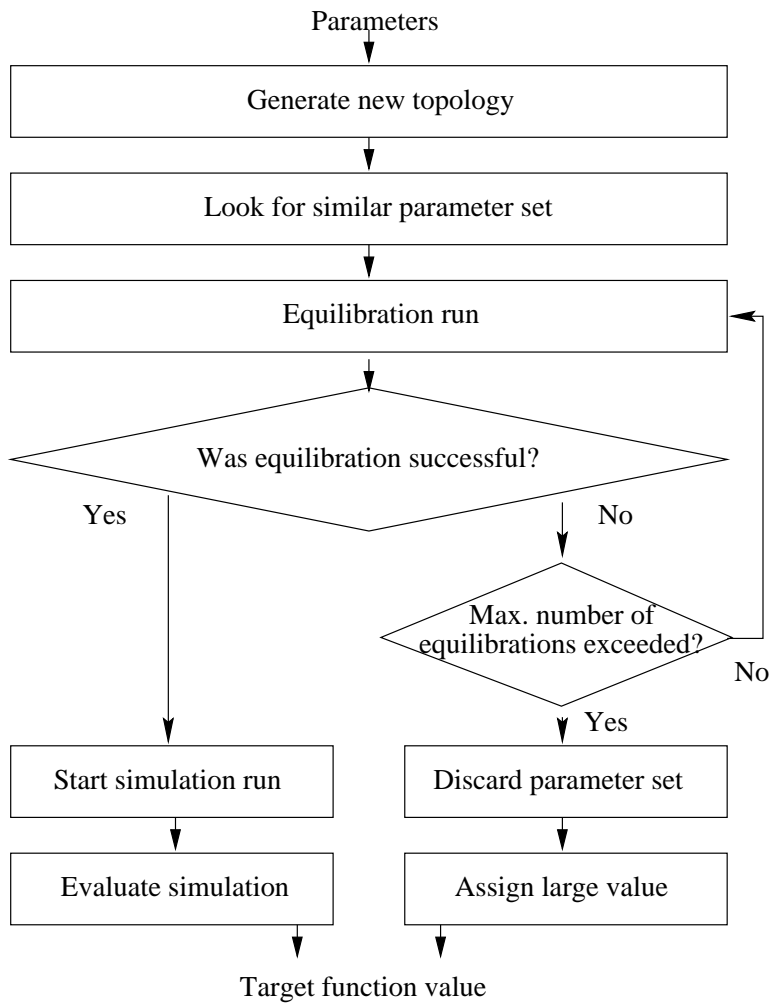}
  \caption{Flow diagram of the algorithm, one iteration.}
  \label{fig:flowdiag}
\end{figure}

The structure of the procedure for obtaining a function value from a
given set of parameters is shown in the flow diagram in figure
\ref{fig:flowdiag}. 
\section{Examples}
The optimization procedure was tested with different model systems in order
to explore its ability to produce force fields.

The all-atom nonbonded force field consists of a Lennard-Jones 12-6 potential
and an electrostatic potential using reaction field and a finite cutoff (of
0.9nm)
\begin{equation}
  V^{(nonb)}=4\epsilon_{ij}\left(\left(\frac{\sigma_{ij}}{r}\right)^{12}-
    \left(\frac{\sigma_{ij}}{r}\right)^{6}\right)+
  \frac{q_{i}q_{j}}{4\pi\varepsilon_{0}\varepsilon}
  \left(\frac{1}{r}+\frac{\varepsilon_{RF}-1}{2\varepsilon_{RF}+1}
    \frac{r^{2}}{r_{cutoff}^{3}}\right). 
\end{equation}
This potential is applied to atoms belonging to different molecules, internal
non-bonded interactions are excluded in our test cases. The Lennard-Jones
parameters between unlike atoms are derived by the Lorenz-Berthelot mixing
rules \cite{allen87}
\begin{equation}
  \epsilon_{ij}=(\epsilon_{ii}\epsilon_{jj})^{\frac{1}{2}}, \quad
  \sigma_{ij}=\frac{1}{2}(\sigma_{ii}+\sigma_{jj}).
\end{equation}
A bond angle potential
\begin{equation}
  V^{(angle)}=\frac{k^{(angle)}}{2}(\Theta-\Theta_{0})^{2},\quad \Theta:
  \mbox{ bond angle}
\end{equation}
and, for some molecules, torsional potentials with threefold symmetry
\begin{equation}
  V^{(tors)}=\frac{k^{(tors)}}{2}\big(1-\cos(3\tau)\big), \quad \tau:
  \mbox{ dihedral angle}
\end{equation}
or a harmonic dihedral potential 
\begin{equation}
  V^{(hd)}=\frac{k^{(hd)}}{2}(\tau-\tau_{0})^{2}
\end{equation}
are applied in order to keep the correct molecular shape.

The bond lengths were constrained using the SHAKE algorithm
\cite{ryckaert77,mplathe91}. Our systems are subject to cubic periodic
boundary conditions. The simulations were run at ambient conditions
(T=$298\,$K, p=$1013\,$hPa). The neighbor-list \cite{allen87} is calculated up
to 1.0nm every 10 to 15 time-steps. We use the Berendsen algorithm for
constant pressure and temperature \cite{berendsen84}. The coupling times were
$0.2\,$ps and 2$\,$ps, respectively. The simulation runs lasted 50$\,$ps at a
timestep of 1$\,$fs for each equilibration run and 100$\,$ps at a timestep of
2$\,$fs for the evaluation runs. The errors of the properties were obtained by
a binning analysis \cite{allen87}.
\subsection{Methylpentane}
As a first test, a system of 125 uncharged 2-methylpentane molecules was
optimized with respect to density $\rho$ and enthalpy of vaporization $\Delta
H_{vap}$.  All Lennard-Jones parameters are subject to optimization. However,
all like atoms (C and H) are constrained to have the same LJ parameters. The
internal part of the force field is taken from the AMBER force field
\cite{amber95}. This comprises the rule that the Lennard Jones $\epsilon$
(eq. \ref{eq:LJ}) is scaled by a factor of 0.5 for 1-4 interactions.

We used the following parameter file to start the algorithm. The number ``4'' 
in the second line indicates the dimensionality of the parameter space. The
following five lines are the guesses of the parameters, the initial
simplex. The last column shows the results after evaluation of the target
function. The Lennard-Jones energies $\epsilon$ and $\Delta H_{vap}$ are
measured in kJ/mol the radii $\sigma$ in nm, the density $\rho$ in kg/m$^{3}$.

\begin{figure}
  \begin{center}
  \epsfysize 8cm
  \epsfbox{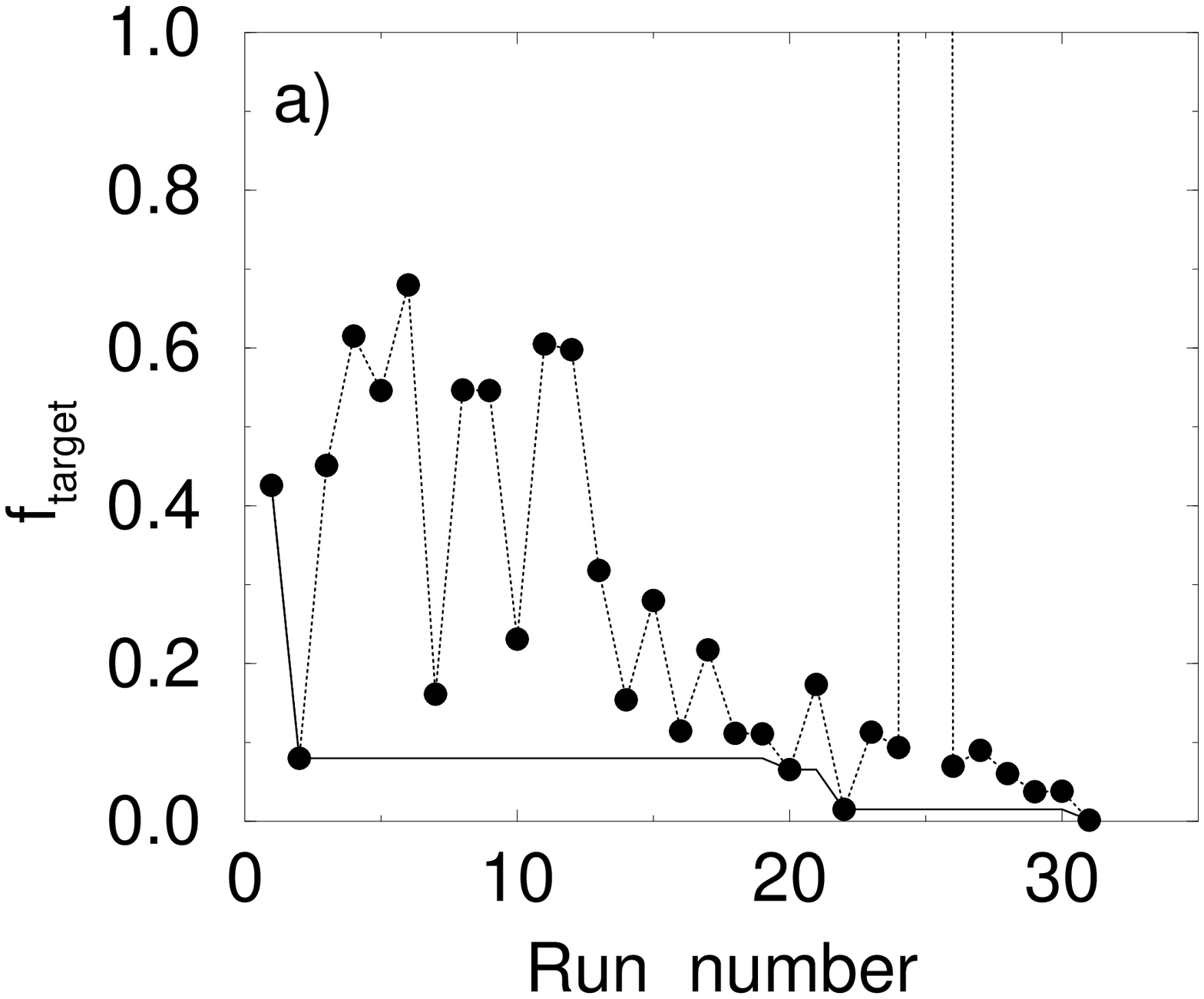}
  \epsfysize 8cm
  \epsfbox{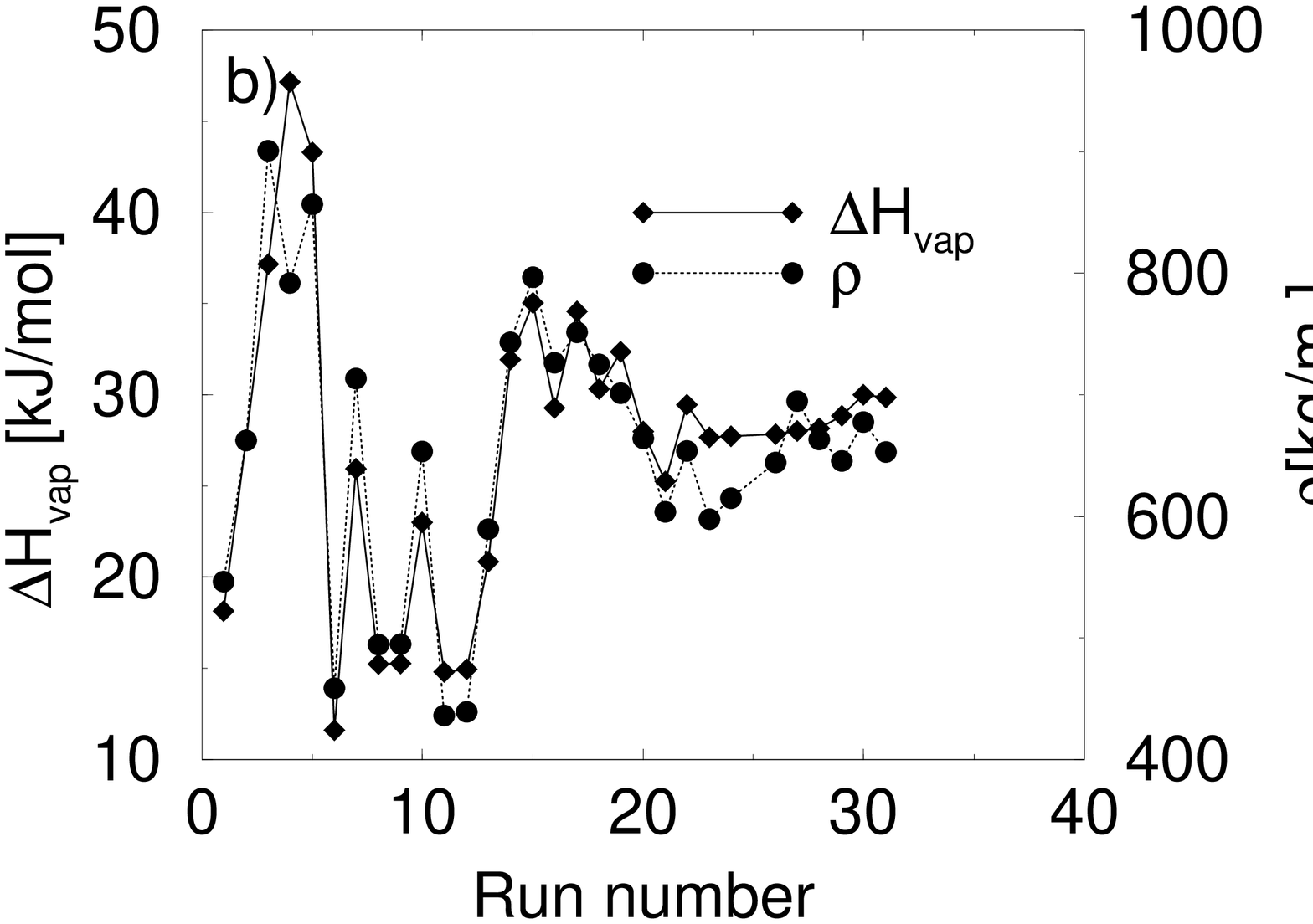}
  \caption{ Convergence of a previous methylpentane optimization run:
a) Target function: Solid line: best value of $f_{target}$; Circles/dotted
line: actual value of $f_{target}$. b) Properties: density and enthalpy of
vaporization.}
  \label{fig:converge}
  \end{center}
\end{figure}

\begin{alltt}
##  \(\epsilon\sb{C}\)      \(\sigma\sb{C}\)     \(\epsilon\sb{H}\)        \(\sigma\sb{H}\)     \(f\sb{target}\)  \(\Delta\)\(H\sb{vap}\)   \(\rho\)
4
0.291643 0.339215 0.154545 0.258859 0.030287 29.42 636.1
0.290554 0.340351 0.151371 0.260571 0.052871 28.88 626.5
0.290167 0.340656 0.151116 0.260825 0.057492 28.81 623.8
0.290545 0.341183 0.149968 0.260762 0.043120 29.18 629.6
0.290421 0.341161 0.150191 0.260763 0.051831 28.94 626.2
\end{alltt}
These parameters produce properties which are already quite close to the
target values. The simplex algorithm now does the fine tuning. First, the
simplex is reflected away from parameter set 3. The new set is
\begin{alltt}
##  \(\epsilon\sb{C}\)      \(\sigma\sb{C}\)     \(\epsilon\sb{H}\)        \(\sigma\sb{H}\)     \(f\sb{target}\)  \(\Delta\)\(H\sb{vap}\)   \(\rho\)
0.291414 0.340299 0.151922 0.259653 0.045721 29.04 629.6
\end{alltt}
After 11 optimization steps, which took about 2 weeks altogether on a DEC
433MHz processor, the optimization finally finished with the following values:
\begin{alltt}
##  \(\epsilon\sb{C}\)      \(\sigma\sb{C}\)     \(\epsilon\sb{H}\)        \(\sigma\sb{H}\)     \(f\sb{target}\)  \(\Delta\)\(H\sb{vap}\)   \(\rho\)
0.294477 0.336339 0.162144 0.254668 0.008629 30.15 652.7
\end{alltt}
Figure \ref{fig:converge} illustrates the progress of the optimization by
means of a previous run. The circles in fig. \ref{fig:converge}a) show
the results of function evaluations and the solid line shows the current best
values of $f_{target}$. In the beginning, the function values scatter quite
strongly. In the run of the algorithm, this starts to decrease.  Figure
\ref{fig:converge}b) shows how density and enthalpy of vaporization reach
their target values.  The only maintenance which had to be done was restarting
the algorithm after a shutdown of the computer system. The whole algorithm
proved to be stable and worked fully automatically. Only once the
equilibration failed due to exceeding the limit of 10 runs. It is shown by the
spike in figure \ref{fig:converge} (which goes up to 100000).  The final force
field is shown in table \ref{tab:metpara}. These values reproduced the
experimental data in a satisfactory way (table \ref{tab:2mpt}).

\subsection{Tetrahydrofurane}
As a different test system we used tetrahydrofurane (THF). Here, we especially
focused on the optimization of partial charges. The hydrogens did not carry
any partial charges but oxygen and carbon did. The charges of the carbons 2
and 5 and the carbons 3 and 4 are the same for symmetry reasons. With the
constraint of electroneutrality, there were two charge parameters to be
optimized. We chose $q_{\mbox{O}}$ and $q_{\mbox{C2/C5}}$, then we have
$q_{\mbox{C3/C4}}=\frac{1}{2}(-q_{\mbox{O}}-2q_{\mbox{C2/C5}})$. Additionally,
the oxygen parameters $\epsilon_{\mbox{O}}$ and $\sigma_{\mbox{O}}$ were
included in the optimization. The first guess for the partial charges was
taken from a quantum chemical Hartree-Fock calculation with a 6-311G** basis
set using Gaussian 94 (Mulliken charges with hydrogens summed into heavy
atoms) \cite{gaussian94}. This yielded also the bond angle values. The bond
lengths are taken from electron diffraction \cite{CRC}. The simulated system
contained 216 molecules. The electrostatic interactions were simulated with a
reaction field correction $(\epsilon_{RF}=7.5)$ using the same cutoff
$r_{c}=0.9\,\mbox{nm}$ as for the Lennard-Jones potential. Here the following
starting simplex was taken:
\begin{alltt}
## \(-q\sb{O}\)   \(q\sb{C2/C5}\)     \(\epsilon\sb{O}\)        \(\sigma\sb{O}\)      \(f\sb{target}\)   \(\Delta\)\(H\sb{vap}\) \(\rho\)
4 
0.581241 0.225443 0.516818 0.208594 0.084176 32.72 816.96
0.658970 0.251733 0.325788 0.300391 0.137257 35.39 811.81
0.480765 0.251793 0.635725 0.316797 0.186735 27.68 774.02
0.684265 0.276431 0.729962 0.264345 0.232852 39.29 847.92
0.582970 0.220928 0.535152 0.192715 0.088840 33.57 823.39
\end{alltt}
The first optimization attempt, which tried to optimize procedure the above
parameters, ended up in a local minimum with $f_{target} \approx 0.07$ after
53 evaluations because the experimental liquid density could not be reproduced
satisfactorily. It was systematically too low. Therefore, the best parameters
so far were frozen and a new optimization was started where only the
Lennard-Jones radii of all species were optimized. Finally, convergence
$(f_{target}\le 0.01)$ was achieved. The resulting THF force field is
described in table \ref{tab:thfforce}.

These parameters lead to the physical properties shown in table
\ref{tab:thfprop}. Our force field has about the same accuracy as an earlier
Monte Carlo simulation of a united atom OPLS model for THF \cite{briggs90}.
\subsection{Cyclic Hydrocarbons}
Finally, the method was applied in order to obtain force fields for
cyclohexene and cyclohexane with 125 molecules in the periodic box. The
geometries were taken from electron diffraction data \cite{CRC}. The geometric
data are shown in table \ref{tab:cygeom}. In the cyclohexene force field,
harmonic dihedral angles are used in order to keep the atoms around the double
bond in plane, since the sp$^{2}$ hybridisation prevents the double bond from
rotating. Additionally, standard torsional potentials with three-fold symmetry
are used. Cyclohexane was simulated without any dihedral angle potentials.
For the angular force constants we used a standard value, since they are
believed to be of minor importance for the desired properties. Additionally,
they may be compensated by the nonbonded parameters.

The optimized Lennard-Jones 12-6 parameters are shown in table
\ref{tab:cypara}. The parameters included in the optimization procedure are
denoted with {\it opt} in the table. No charges were used. All the parameters
which are not optimized as well as the initial simplices have been taken from
similar force fields. 

Except for the Lennard-Jones $\epsilon$ of the hydrogens, the resulting final
parameters are very similar for the two molecules. This shows that force field 
parameters are not a unique description of a certain atom type but rather they 
are only a part of the overall molecular description. Mostly, however, the
same atoms in similar environments may be described by similar parameters.

We compare our thermodynamic data with experiment in table \ref{tab:cyprop}. A
more detailed analysis of transport properties of these cyclic hydrocarbons
will be published elsewhere \cite{schmitz99sa}. The cyclohexane force field
yields a slightly better comparison to experiment than in a recent study using
a commercial force field \cite{sun98a} whereas the study of cyclohexene is the
first to our knowledge. 
\section{Conclusions}
We applied the simplex algorithm to the problem of force field optimization
for MD simulations. Given a good initial guess for the force field parameters
and the experimental data for some properties, our method tunes the parameters
to optimum values. Once the routine has been set up, very little human
interference is required for maintenance. The algorithm proved to be robust
and found local minima if set up properly. The resulting force fields are able
to reproduce experimental data of low molecular weight liquids in a
satisfactorily. 

%for referees
In the examples of this contribution, we typically optimized 4 force field
parameters against 2 observables. Hence, the solutions are most likely not
unique. This, however, is a feature of the problem of finding a force field
given a small number of observables, not of the algorithmic solution presented
here. Density and enthalpy of vaporization are the two properties most
commonly used to derive force fields, as they are experimentally available for
many fluids and quickly converging in a simulation. At present, our method has
to be used with a judicious choice of starting values for the parameters to
prevent it from optimizing towards an unphysical, non-transferable set of
parameters. It shares this restriction with all other methods of finding force
fields, including ``optimization by hand''. On the other hand, it is mostly
not difficult to come up with a reasonable first guess for the
parameters. What is time consuming is the fine tuning and it is at this point
where our method offers help. 

A possible way out of the dilemma is to increase the base of experimental
observables used in the target function. In a few selected cases we have used
other liquid properties than $\rho$ and $\Delta H_{vap}$ together with the
other refinement scheme \cite{njo95,berweger95}. However, one has to note that
there are not too many suitable fluid properties. Some properties are of
similar character to what we already have. For example, the excess chemical
potential $\mu_{ex}$ probes almost the same regions of the force field as
$\Delta H_{vap}$ and, thus, does not add much independent information. Dynamic
properties often converge too slowly in simulations to be useful (shear
viscosity, dielectric constant) or the experimental data are not of sufficient
quality (tracer diffusion coefficient, molecular reorientation times). We,
therefore, follow the strategy of optimizing towards $\rho$ and $\Delta
H_{vap}$ and subsequently checking the final force field against other liquid
properties. For our models of cyclic hydrocarbons we have, for instance,
calculated tracer and binary diffusion coefficients as well as molecular
reorientation times for both the pure liquids and binary mixtures, and the
results agree well with experimental data where available \cite{schmitz99sa}.
%ende

The automatic parameterization scheme presented has the small disadvantage of
probably requiring moderately more computer time than an optimization by
hand. This is more than offset by the invaluable advantage of freeing
researchers from the labour of parameter optimization.
% sentence added for referee
In a reasonable use of computing time (a few weeks workstation time) one is
able to cope with dimensionalities of parameter space of about 4. This
depends, however, strongly on the actual simulations to be performed. On the
other hand, the full potential of speeding up our algorithm has not yet been
realized. We foresee possibilities of substantial improvement by using a less
rigorous and maybe adaptive equilibration scheme and by substituting the
simplex algorithm by a faster converging optimizer (e.g. Fletcher) in the
final stages of minimization. This remains an interesting starting point for
future research.

\bibliographystyle{jcc}
\bibliography{standard}

\pagebreak

\pagestyle{empty}

\begin{table}
  \begin{center}
    \begin{tabular}[t]{|c|c||c|c|}
      \hline
      \multicolumn{2}{|c||}{bonded parameters} & 
      \multicolumn{2}{|c|}{non-bonded parameters}\\
      \hline
      parameter & value & parameter & value\\
      \hline
      $|$C-C$|$ & 0.1526$\,$nm & $m_{C}$ & 12.01$\,$amu \\
      $|$C-H$|$ & 0.109$\,$nm & $m_{H}$ & 1.00782$\,$amu \\
      $k^{(angle)}_{C-C-C}$ & 167.47\angen& $\epsilon_{C}$ & 0.294$\,$kJ/mol\\
      $k^{(angle)}_{C-C-H}$ & 209.34\angen& $\epsilon_{H}$ & 0.162$\,$kJ/mol\\
      $k^{(angle)}_{H-C-H}$ & 146.54\angen& $\sigma_{C}$ & 0.336$\,$nm\\
      H-C-H & 109.5$^{\circ}$ & $\sigma_{H}$ & 0.254$\,$nm\\
      C-C-C & 109.5$^{\circ}$ & &\\
      C-C-H & 109.5$^{\circ}$ & &\\
      $k^{(tors)}_{C-C-C-C}$ & 11.5kJ/mol & &\\
      $k^{(tors)}_{methyl}$  & 11.5kJ/mol & &\\
      \hline
    \end{tabular}
    \caption{Details of the Methylpentane force field}
    \label{tab:metpara}
  \end{center}
\end{table}

%\pagebreak\clearpage

\begin{table}
  \begin{center}
    \begin{tabular}{|c|c|c|}
      \hline
      & exp. & sim. \\
      \hline
      $\Delta H_{vap}$[kJ/mol] & 29.89\cite{CRC} & 29.92$\pm$0.03\\
      $\rho$[kg/m$^{3}$] & 653.0\cite{CRC} & 653.4$\pm$0.5\\
      $D$[cm$^{2}$/s] & & (2.5$\pm$0.2)$\times$10$^{-5}$\\
      \hline
    \end{tabular}
    \caption{Experimental and simulated properties of 2-methylpentane}
    \label{tab:2mpt}
  \end{center}
\end{table}

%\pagebreak\clearpage

\begin{table}
  \begin{center}
    \begin{tabular}[t]{|c|r||c|r|}
      \hline
      \multicolumn{2}{|c||}{nonbonded parameters} & 
      \multicolumn{2}{|c|}{bonded parameters}\\
      \hline
      parameter & \multicolumn{1}{c||}{value} & parameter & 
      \multicolumn{1}{c|}{value}\\
      \hline
      $\epsilon_{\mbox{O}}$ & 0.509$\,$kJ/mol & $|$C-O$|$ & 0.1428$\,$nm\\
      $\epsilon_{\mbox{H}}$ & 0.200$\,$kJ/mol & $|$C-H$|$ & 0.1115$\,$nm\\
      $\epsilon_{\mbox{C}}$ & 0.290$\,$kJ/mol & $|$C-C$|$ & 0.1536$\,$nm\\
      $\sigma_{\mbox{O}}$ & 0.243$\,$nm & $k^{(angle)}$ & 450.0\angen\\
      $\sigma_{\mbox{H}}$ & 0.193$\,$nm & C-O-C & 111.2$^{\circ}$\\
      $\sigma_{\mbox{C}}$ & 0.306$\,$nm & O-C-C & 106.1$^{\circ}$\\
      $q_{\mbox{O}}$ & $-0.577\,e$ & C-C-C & 101.4$^{\circ}$\\
      $q_{\mbox{C2}}$ & $0.228\,e$ & O-C-H & 109.0$^{\circ}$, 
      109.3$^{\circ}$\\
      $q_{\mbox{C3}}$ & 0.061$\,e$ & C$_{3}$-C$_{2}$-H & 111.0$^{\circ}$, 
      113.2$^{\circ}$\\
      $m_{\mbox{O}}$ & 15.9949$\,$u & H-C$_{2}$-H & 108.2$^{\circ}$\\
      $m_{\mbox{C}}$ & 12.0$\,$u & H-C$_{3}$-H & 108.1$^{\circ}$\\
      $m_{\mbox{H}}$ & 1.00787$\,$u & C$_{2}$-C$_{3}$-H & 110.4$^{\circ}$ 
      (2$\times$)\\
      & & & 112.8$^{\circ}$(2$\times$) \\
      & & C$_{3}$-C$_{4}$-H & 113.7$^{\circ}$ (2$\times$)\\
      & & & 110.4$^{\circ}$ (2$\times$)\\
      \hline
    \end{tabular}
    \caption{Optimized force field for tetrahydrofurane. In the case of two
      angles in one line one is applied to the first hydrogen, the other to
      the second hydrogen, otherwise the angles would not be consistent with
      each other. }
    \label{tab:thfforce}
  \end{center}
\end{table}

%\pagebreak\clearpage

\begin{table}
  \begin{center}
    \begin{tabular}{|c|c|c|c|}
      \hline
      & experiment\cite{CRC} & simulation (this work) & 
      simulation \cite{briggs90}\\
      \hline
      $\rho$ &        889.0$\,$kg/m$^{3}$ & (886.0 $\pm$ 
      1.3)$\,$kg/m$^{3}$ & (882$\pm$1)$\,$kg/m$^{3}$\\
      $\Delta H_{vap}$ & 31.99$\,$kJ/mol & 
      (32.0$\pm$0.1)$\,$kJ/mol & (31.57$\pm$0.08)$\,$kJ/mol\\
      \hline
    \end{tabular}
    \caption{Properties of tetrahydrofurane} 
    \label{tab:thfprop}
  \end{center}
\end{table}

%\pagebreak\clearpage

\begin{table}
  \begin{center}
    \begin{tabular}[t]{|c|cc|}
      \hline
      property & C$_{6}$H$_{10}$ & C$_{6}$H$_{12}$  \\
      \hline
      $|$C$_{sp2}$=C$_{sp2}|$ & 0.1334nm & \\
      $|$C$_{sp2}$-C$_{sp3}|$ & 0.150nm & \\
      $|$C$_{3}$-C$_{4}|$, $|$C$_{5}$-C$_{6}|$ & 0.152nm & \\
      $|$C$_{4}$-C$_{5}|$ & 0.154nm& \\
      $|$C-C$|$ & & 0.1526nm \\
      $|$C$_{sp2}$-H$|$ & 0.108nm & \\
      $|$C$_{sp3}$-H$|$ & \multicolumn{2}{c|}{0.109nm}\\
      $k^{(angle)}_{\mbox{C-C-C}}$ & 450\angen & 335\angen \\
      $k^{(angle)}_{\mbox{C=C-C}}$ & 500\angen & \\
      $k^{(angle)}_{\mbox{C-C-H}}$ & 500\angen & 420\angen\\
      $k^{(angle)}_{\mbox{H-C-H}}$ & 500\angen & 290\angen\\
      C=C-C & 112.0$^{\circ}$ & \\
      C-C-C & 110.9$^{\circ}$ & 109.5$^{\circ}$\\
      C$_{sp2}$-C-C & 123.45$^{\circ}$ & \\
      C-C-H & \multicolumn{2}{c|}{109.5$^{\circ}$}\\
      C-C$_{sp2}$-H & 119.75$^{\circ}$ & \\
      H-C-H & \multicolumn{2}{c|}{109.5$^{\circ}$}\\
      $k^{(hd)}_{\mbox{C-C=C-C}}$ & 250\angen & \\
      $k^{(hd)}_{\mbox{H-C=C-C}}$ & 200\angen & \\
      $k^{(tors)}_{\mbox{C-C-C-C}}$ & 10kJ/mol & \\
      \hline
    \end{tabular}
    \caption{Geometry of the cyclic hydrocarbons and their intramolecular
      potentials}
      \label{tab:cygeom}
  \end{center}
\end{table}

%\pagebreak\clearpage

\begin{table}
  \begin{center}
    \begin{tabular}[t]{|l|l|cc|}
      \hline
      parameter & opt/fix & C$_{6}$H$_{10}$ &C$_{6}$H$_{12}$\\
      \hline
      $\epsilon_{\mbox{C}}$ & opt & 0.296kJ/mol & 0.299kJ/mol\\
      $\epsilon_{\mbox{H}}$ & opt & 0.265kJ/mol & 0.189kJ/mol\\
      $\sigma_{\mbox{H}}$ & opt & 0.252nm &  0.258nm\\
      $\sigma_{\mbox{C}}$ & opt &  & 0.328nm\\
      $\sigma_{\mbox{C}sp2}$ & fix & 0.321nm &  \\
      $\sigma_{\mbox{C}sp3}$ & fix & 0.311nm &  \\
      $m_{\mbox{C}}$ & fix & \multicolumn{2}{c|}{12.01amu}\\
      $m_{\mbox{H}}$ & fix & \multicolumn{2}{c|}{1.00787amu} \\
      \hline
    \end{tabular}
    \caption{Cyclohexene and cyclohexane non-bonded parameters}
    \label{tab:cypara}
  \end{center}
\end{table}

%\pagebreak\clearpage

\begin{table}
  \begin{center}
    \begin{tabular}{|l|c|c|c|c|c|}
      \hline
      \multicolumn{1}{|c|}{}&\multicolumn{2}{c|}{cyclohexene} &
      \multicolumn{3}{c|}{cyclohexane}\\
      \hline
      & exp & sim & exp\cite{CRC} & sim (this work)& sim \cite{sun98a}\\
      \hline
      $\rho$ [kg/m$^{3}$]& 805.8 \cite{harris94} & 806.0 $\pm$ 1.5 &
       777.6  & 775.9$\pm$0.8 & 774$\pm$2\\
      $\Delta H_{vap}$ [kJ/mol] & 33.47 \cite{CRC} & 33.3 $\pm$ 0.1 &
      33.33 & 33.46$\pm$0.05 & 33.41\\
      \hline
    \end{tabular}
  \end{center}
  \caption{Properties of cyclohexene and cyclohexane}
  \label{tab:cyprop}
\end{table}

%\pagebreak\clearpage

%\noindent Figure 1: Transformations of the simplex used during the algorithm: 
%a) reflection, b) expansion, c) contraction.

%\vspace{2cm}

%\noindent Figure 2: Flow diagram of the algorithm, one iteration.

%\vspace{2cm}

%\noindent Figure 3: Convergence of a previous methylpentane optimization run:
%a) Target function: Solid line: best value of $f_{target}$; Circles/dotted
%line: actual value of $f_{target}$. b) Properties: density and enthalpy of
%vaporization.

%\pagebreak\clearpage

%\pagebreak\clearpage

%\pagebreak\clearpage

\end{document}